\documentclass[aps,prlprint, twocolumn,superscriptaddress,amsmath,amssymb,showpacs]{revtex4}

\usepackage{graphicx}
\begin{document}
\title{\textbf{Enhanced laser-driven ion acceleration by superponderomotive electrons generated from near-critical-density plasma}}
\author{J.\,H. Bin}\email[]{jianhuibin@lbl.gov}
\affiliation{Fakult$\ddot{a}$t f$\ddot{u}r$ Physik, Ludwig-Maximilians-Universit$\ddot{a}$t M$\ddot{u}$nchen, D-85748 Garching, Germany}
\affiliation{Max-Planck-Institut f$\ddot{u}$r Quantenoptik, D-85748 Garching, Germany}
\affiliation{Lawrence Berkeley National Laboratory, University of California, Berkeley, California 94720, USA}

\author{M. Yeung}
\affiliation{Department of Physics and Astronomy, Centre for Plasma Physics, Queens University Belfast, BT7 1NN, UK}

\author{Z. Gong}
\affiliation{State Key Laboratory of Nuclear Physics and Technology, and Key Laboratory of High Energy Density Physics Simulation, CAPT, Peking University, 100871 Beijing, China}

\author{H.Y. Wang}
\affiliation{Helmholtz Institute Jena, Fr$\ddot{o}$belstieg 3, 07443 Jena, Germany}

\author{C. Kreuzer}
\affiliation{Fakult$\ddot{a}$t f$\ddot{u}r$ Physik, Ludwig-Maximilians-Universit$\ddot{a}$t M$\ddot{u}$nchen, D-85748 Garching, Germany}

\author{M.L. Zhou}
\affiliation{State Key Laboratory of Nuclear Physics and Technology, and Key Laboratory of High Energy Density Physics Simulation, CAPT, Peking University, 100871 Beijing, China}



\author{M.J.V. Streeter}
\affiliation{Blackett Laboratory, Imperial Colleage London, SW7 2BZ, UK}

\author{P.S. Foster}
\affiliation{Department of Physics and Astronomy, Centre for Plasma Physics, Queens University Belfast, BT7 1NN, UK}
\affiliation{Central Laser Facility, STFC Rutherford Appleton Laboratory, Chilton, Didcot, Oxon, 0X11 0QX, UK}

\author{S. Cousens}
\affiliation{Department of Physics and Astronomy, Centre for Plasma Physics, Queens University Belfast, BT7 1NN, UK}


\author{B. Dromey}
\affiliation{Department of Physics and Astronomy, Centre for Plasma Physics, Queens University Belfast, BT7 1NN, UK}

\author{J. Meyer-ter-Vehn}
\affiliation{Max-Planck-Institut f$\ddot{u}$r Quantenoptik, D-85748 Garching, Germany}

\author{M. Zepf}
\affiliation{Department of Physics and Astronomy, Centre for Plasma Physics, Queens University Belfast, BT7 1NN, UK}
\affiliation{Helmholtz Institute Jena, Fr$\ddot{o}$belstieg 3, 07443 Jena, Germany}

\author{J. Schreiber}\email[]{Joerg.Schreiber@lmu.de}
\affiliation{Fakult$\ddot{a}$t f$\ddot{u}r$ Physik, Ludwig-Maximilians-Universit$\ddot{a}$t M$\ddot{u}$nchen, D-85748 Garching, Germany}
\affiliation{Max-Planck-Institut f$\ddot{u}$r Quantenoptik, D-85748 Garching, Germany}

\date{\today}

\begin{abstract}
 
We report on the experimental studies of laser driven ion acceleration from double-layer target where a near-critical density target with a few-micron thickness is coated in front of a nanometer thin diamond-like carbon foil. A significant enhancement of proton maximum energies from 12 to $\sim$30 MeV is observed when relativistic laser pulse impinge on the double-layer target under linear polarization. We attributed the enhanced acceleration to superponderomotive electrons that were simultaneously measured in the experiments with energies far beyond the free-electron ponderomotive limit. Our interpretation is supported by two-dimensional simulation results.

\end{abstract}
 \maketitle
 
The rapid development of high-power laser technology has triggered a fast evolution of laser driven particle sources. In the past two decades, ion beams generated by relativistic laser pulses interacting with solid-density targets have stimulated the idea of compact particle sources for a range of applications, and thus attract great attentions \cite{BorghesiFST2006, DaidoRPP2012, MacchiRMP2013, SchreiberRSI2016}. Investigations have been dedicated to advanced targets such as nanometer thin foil \cite{MaNIMPRS2011, BinPOP2013} and novel acceleration mechanisms such as breakout afterburner (BOA) \cite{YinLPB2006, YinPOP2007} and the light sail form of radiation pressure acceleration (RPA-LS) \cite{MacchiPRL2005, KlimoPRSTAB2008, YanPRL2008, RobinsonNJP2008, QiaoPRL2009}. The latter one is a result of accelerating fields that established when radiation pressure of the incident laser pulse pushing on plasma electrons which drag ions along, promising higher ion energy, well-controlled energy spectrum, and higher conversion efficiency. By suppression of heating the electrons when using normal incidence and circular polarization (CP), RPA-LS has been recently demonstrated in experiments \cite{HenigPRL2009, KarPRL2012, SteinkePRST2013, BinPRL2015}.

Alternatively, it is of great interest to tackle this issue with the most experimental-investigated mechanism, target normal sheath acceleration (TNSA) mechanism \cite{WilksPOP2001}, where ions are accelerated by strong electrostatic fields that are induced when hot electrons cross the target and exit the target. The fact that hot electrons are generated in the target front by absorbing laser energy, triggers enthusiastic prospects of improving acceleration by increase laser absorption and conversion efficiency to hot electrons. For instant, foam-based double-layer target has been proposed by theorists \cite{TatsufumiPOP2010, SgattoniPRE2012, WangPOP2013} and tested in experiments recently \cite{PrencipePPCF2016}. Increased proton acceleration has been reported with targets coated with microspheres monolayer \cite{MargaronePRL2012}. In particular, superponderomotive electrons from laser-plasma interaction, which has been recently revisited both experimentally and thoertically \cite{JiangPRL2016, SorokovikovaPRL2016}, could be another promising candidate to increase ion acceleration, which has not yet been demonstrated in experiments so far.

In this letter, we present a novel approach to enhance ion acceleration based on a double-layer target configuration. In particular, nanometer (nm) thin diamondlike carbon (DLC) foils \cite{MaNIMPRS2011} coated with several $\mu m$ thick carbon nanotube foams (CNF) \cite{MaNL2007} at near-critical-density (NCD) are used in experiments. In contrast to our previous investigating case under circular polarization (CP) \cite{BinPRL2015}, the electron density of CNF employed here is about 0.5$n_{c}$, slightly below the critical density of electrons $n_{c}$. Significant enhancement of ion beams that accelerated from such targets as compared to pure DLC foils, has been measured in the experiments with linearly polarized incident laser pulses. The electron spectrum that was monitored in parallel suggest that the increased acceleration field can be attributed to strongly enhanced electron heating to average energies much beyond the free electron ponderomotive limit. Supporting two-dimensional (2D) particle-in-cell (PIC) simulations reveal that those superponderomotive electrons are generated from the several $\mu m$ thick NCD plasma during the interaction via direct laser acceleration (DLA) mechanism \cite{PukhovPOP1999, GahnPRL1999, ManglesPRL2005}.

The experiments were performed at the Gemini laser at the Central Laser Facility of the Rutherford Appleton Laboratory in the UK. The system delivers pulses with duration of 50 fs full-width at half-maximum (FWHM) centered at 800 nm wavelength. A re-collimating double plasma mirror system was introduced to enhance the laser contrast to a ratio of $10^{-9}$ at 5 ps before the peak of main pulse. 4-5 J laser energy was delivered on target. A f/2 off-axis parabolic mirror (OAP) was used to focus the laser pulses to a FWHM focal spot diameter of 3.5 ${\mu m}$, yielding a peak intensity of $2\times10^{20}$ W/cm$^2$ ($a_{0}\approx$ 10) when considering the complete spatial distribution with high dynamic range. 20 nm DLC foils were coated with CNF layers with various thicknesses and have been irradiated under normal incidence with linear polarization (LP). 

\begin{figure}[!ht]
  \includegraphics[width=1.0\columnwidth]{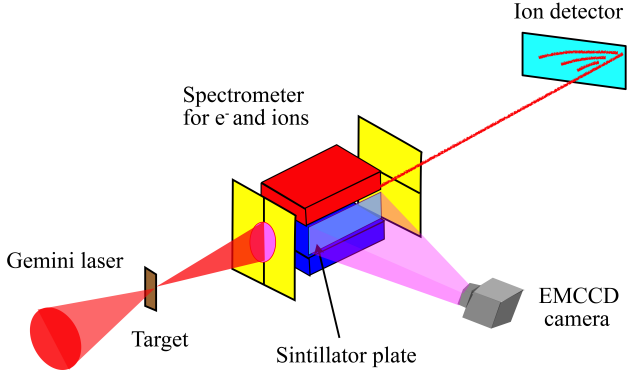}
  \caption{(color online). Experimental setup. The Gemini laser pulse is normally incident on the target under linear polarization. A special designed spectrometer provides simultaneous measurements of electron and ion spectrum.} \label{fig1}
\end{figure}

The experimental setup is sketched in Fig. \ref{fig1}. In the experiments, electrons and ions were simultaneous determined with a special designed spectrometer with moderate b-field strength (100-240 mT). Ions were measured in the target normal with two sequential, mutually perpendicular slit entrances  which acting as a Thomson-parabola-like setup. The electrons were deflected to a scintillator plate located inside the magnets and imaged onto a EMCCD camera.

\begin{figure}[!ht]
  \includegraphics[width=1.0\columnwidth]{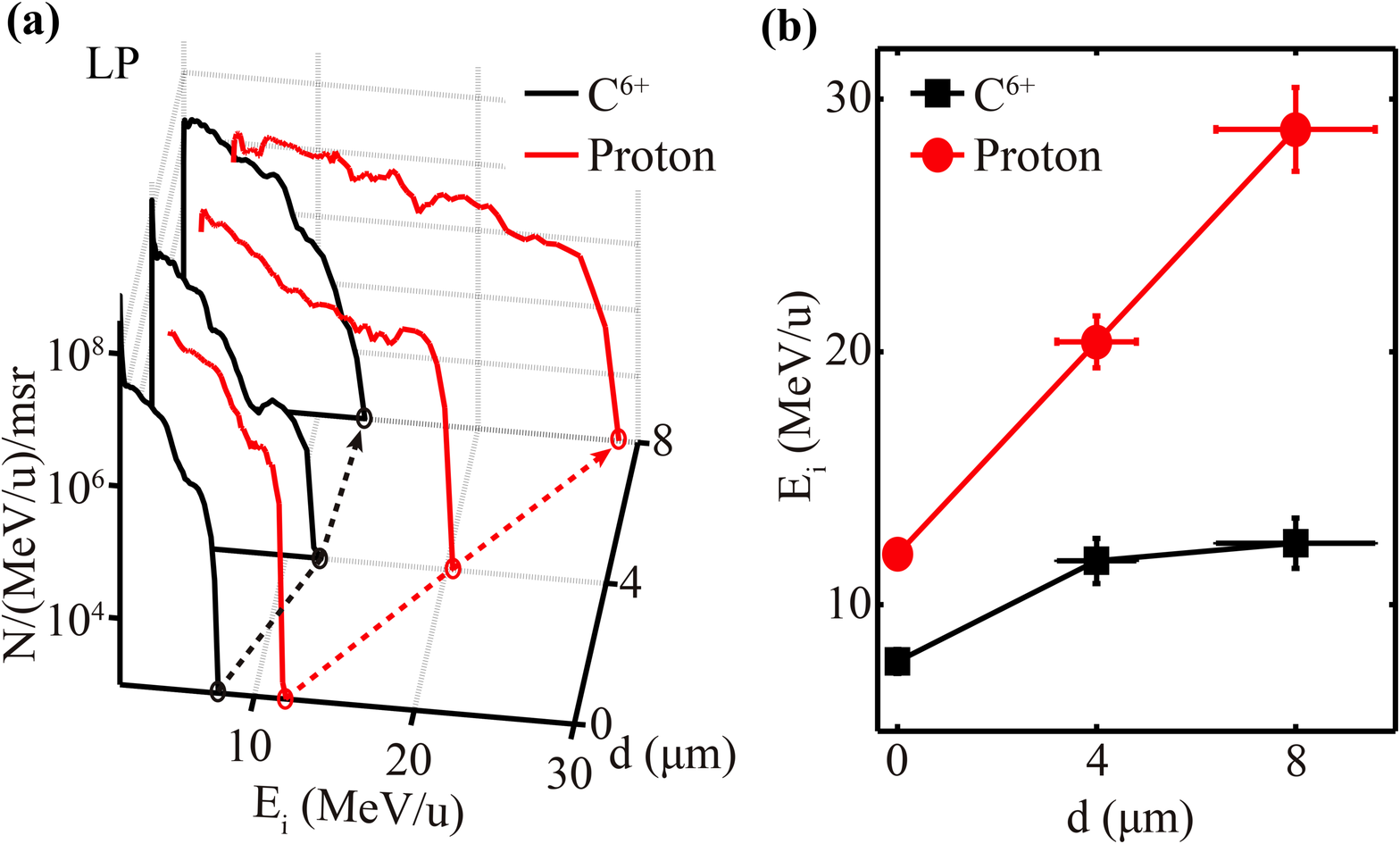}
  \caption{(color online). (a) Ion energy distributions of protons (red) and C$^{6+}$ (black) for the case of linearly polarized laser pulses interacting with CNF of varying thickness attached to a 20 nm DLC foil. (b) Maximum energy of proton (red) and carbon (black) beams for varying CNF thickness in double-layer target configuration. The vertical error bars denote the energy resolution of the spectrometer and the horizontal error bars present the production deviation in target thickness.} \label{fig2}
\end{figure}

The ion results are summarized in Fig. \ref{fig2} for the LP Gemini laser pulses. The energy distributions monotonically decay and terminate at a maximum energy value, both for protons and carbon ions (see Fig. \ref{fig2} (a)). The energies increase with increasing CNF thickness. In particular, the maximum energy of protons increases more strongly, from 12 to 29 MeV - a factor of 2.4 - with increasing thickness of the CNF layer, while C$^{6+}$ energy increases by a smaller factor of 1.7 (Fig. \ref{fig2} (b)). When using freestanding CNF targets, no proton signal and only low energy carbons with maximum energy up to $\sim$ 3.5 MeV/u has been detected.   

\begin{figure}[!ht]
  \includegraphics[width=1.0\columnwidth]{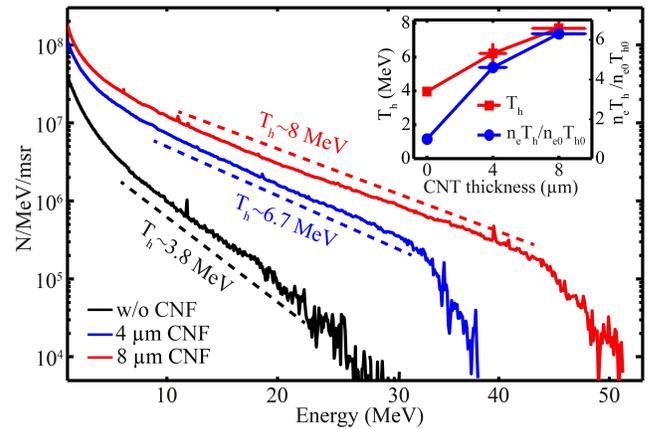}
  \caption{(color online). Measured electron energy spectra along target normal (laser propagating direction) with LP pulses interacting with CNF with different thickness attached to a 20 nm DLC foil. The inset shows the comparison of corresponding electron temperature $T_{h}$, and total electron energy as presented by the production $n_{e}T_{h}$ after normalization to the value $n_{e0}T_{h0}$ from 20 nm DLC foil solely. Both quantities $n_{e}$ and $T_{h}$ are extracted from the fitting curve.} \label{fig3}
\end{figure}

Fig. \ref{fig3} shows the energy spectra of electrons measured simultaneously to the corresponding ion spectra in Fig. \ref{fig2}. The spectrum measured from pure 20 nm DLC foil (w/o CNF) presents a typical quasi-thermal distribution (black curve). An exponential fit to the measured electron energy distribution yields an electron temperature $T_{h}$ of 3.8 MeV, comparable to the ponderomotive temperature $T_{h, pond}=(\sqrt{1+a_{0}^2/2}-1)m_{e}c^2$ \cite{KruerPOF1985} for $a_{0}=10$. 

 \begin{figure*}[!ht]
  \includegraphics[width=2.0\columnwidth]{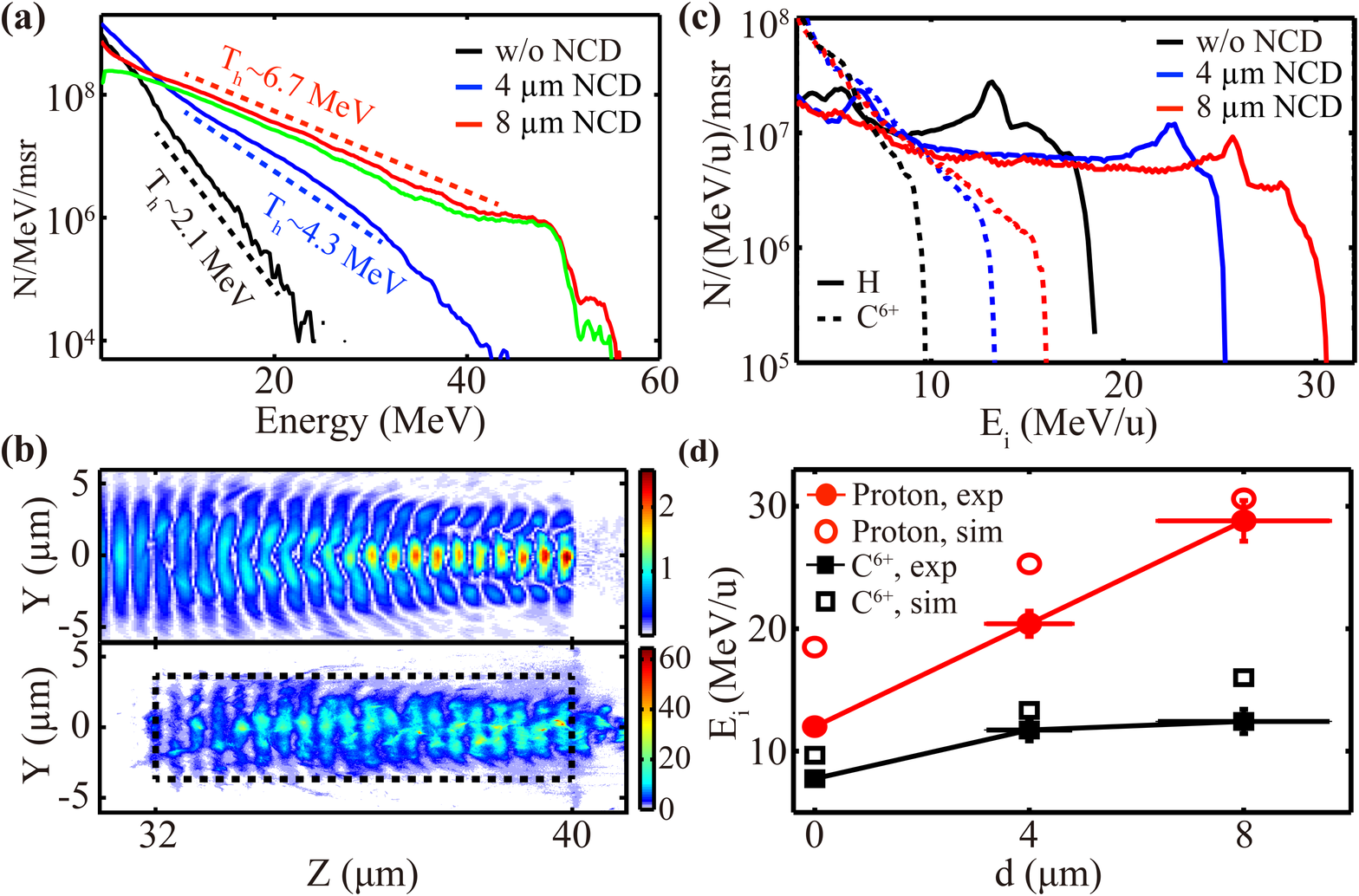}
  \caption{(color online). 2D PIC simulation results. (a) Absolute electron energy distribution with different NCD plasma length, 
  the time is chosen when the electron cut-off energy reaches a maximum in individual cases. The green curve shows the electron spectrum from the area identified by the black dashed box in (b). (b) Simulated laser field distribution normalized by initial $a_{0}$ (upper) and energy density of the electrons from NCD plasma normalized by $n_{c}m_{e}c^2$ (lower) for the 8 $\mu m$ case in (a). Here the NCD plasma is located in the range of 32-40 $\mu m$. The black dashed box marks the DLA electrons from NCD plasma within laser focal volume. (c) C$^{6+}$ ion (dashed curves) and proton (solid curves) energy spectra extracted from simulations at $t=$ 250 fs for varying NCD thickness (color marked). The maximum energies (empty symbols) are compared with the experimental data (solid symbols) in (d).} \label{fig4}
\end{figure*}

With additional CNF layer attached to the DLC foil, more pronounced electron heating is observed. Both the temperature $T_{h}$ and the total energy which is represented by $n_{h}T_{h}$ of hot electrons increase with CNF thickness. The most energetic electron spectrum was obtained with 8 $\mu m$ thick CNF layer, yielding superponderomotive electrons with $T_{h}$ of about 8 MeV, much higher than the free-electron ponderomotive scaling at identical intensity ($T_{h}=$ 3.6 MeV), and the total electron energy $n_{e}T_{h}$ is enhanced by a factor of 6.2 as compared to single layer DLC foil (see the inset in Fig. \ref{fig3}). The vertical error bars denote the difference between estimated $T_{h}$ to $\overline{E}-E_{min}$, where mean electron energy $\overline{E}$ is deduced within the resolved spectral range starting from optional minimum energy $E_{min}=$ 10 MeV. In other words, they present the deviation from the measured spectral to an ideal exponential curve. The horizontal error bars represent the production deviation in target thickness. 

The strong correlation between the spectra of electrons and ions suggests the important role of those superponderomotive electrons in the acceleration process and in fact is consistent with a simple physical picture. The ions are accelerated in an electric field set up by fast laser-accelerated electrons, the accelerating field is thus determined by $E_{acc} \propto \sqrt{n_{e}T_{h}}$ regardless of the actual dynamics \cite{MoraPRL2003, SchreiberPRL2006}. One would then expect the resultant ion energy scale as $\sqrt{n_{e}T_{h}}$. Off course, in practice the exact dependence might be complicated by details of the dynamics. Nevertheless, the observed superponderomotive electrons is expected to enhance laser driven ion acceleration. In fact, our interpretation is supported by our experimental results where we found that the optimal enhancement (2.4$\times$), which suggests that the accelerating field is increased by a factor of 2.4, is consistent with the measured enhancement in total energy of superponderomotive electrons $n_{e}T_{h}$ (6.2$\times$). The reduced energy gain of carbon (1.7$\times$) suggests that they experience the same acceleration field integral but with only half the charge to mass ratio compared to protons.

To further understand the interaction, two-dimensional (2D) particle-in-cell (PIC) simulations were carried out with EPOCH code \cite{ArberPPCF2015}. Solid density (80 nm, $n_0=150n_c$, where $n_c$ is the critical density) plasma slab was used to present the employed DLC foils. NCD plasma slab (0.5$n_{c}$) with different lengths represent the CNF targets. The initial temperature of electrons is 1 keV. The simulation box is 100 $\lambda$ in laser direction ($Z$) and 30 $\lambda$ in transverse direction ($Y$) in 2D with a resolution of 100 cells/$\lambda$ and 25 cells/$\lambda$, respectively. Each cell is filled with 28 macroparticles. A linearly polarized laser pulse, with peaked intensity $I_{0}=2\times10^{20}$ W/cm$^2$, with a Gaussian envelope in both the spatial  and temporal distribution with a FWHM diameter  $D_{L}$ of 3.5 $\mathrm{\mu m}$ and a FWHM duration of 50 fs, is used to approximate the experiment conditions. 
 
Fig. \ref{fig4} (a) shows the quantitative energy distribution of electrons obtained from double-layer targets with different NCD plasma length. We observe a significant increase with increasing NCD plasma length, agreeing well with our experimental results. In particular, the superponderomotive behavior is reproduced with thicker NCD plasma. The simulation result indicates an efficient DLA acceleration process when the laser propagates through the NCD plasma (upper graph of Fig. \ref{fig4} (b)), the electron energy density distribution closely resemble the laser intensity distribution including a clear resonance behavior (lower graph of Fig. \ref{fig4} (b)). This leads to superponderomotive electrons with both higher electron temperature and electron number. In fact, these DLA electrons constitutes the major part to the total electron energy distribution, as indicated by the green curve in Fig. \ref{fig4} (a) which is extracted from the marked area in black dashed box. Note that, relativistic self-focusing \cite{BorghesiPRL1997} can benefit the generation of superponderomotive electrons \cite{WangPOP2013}. While, as we only employed NCD targets with much lower density, the target thickness in our case is far below the optimum length for relativistic self-focusing effect. For example, one would expect a self-focusing length of about 16 $\mu m$ with NCD target at density of 0.5$n_c$ \cite{BinPRL2015}. Therefore we judge the impact from relativistic self-focusing under our condition small. It could be further explored in future experiments. 

As expected, the superponderomotive electrons give rise to enhanced ion acceleration, as evidenced in Fig. \ref{fig4} (c) and (d). In agreement with the experiments, we observed strong enhancement of both proton and carbon ion energies with increasing NCD plasma length. In general, the energy distributions for both species exhibit monotonically decaying shape, similar to experimental observations. More important, the observed energy per nucleon of protons are by a factor of $\sqrt{2}$ larger than carbon ions, a feature that is expected from plasma expansion mechanism \cite{MoraPRL2003, SchreiberPRL2006}. Further supports on our hypothesis is shown in Fig. \ref{fig4} (d), where the maximum energies were extracted from simulations and plotted versus the NCD thickness, showing fair agreement with the experimental results.

In conclusion, we have demonstrated the positive effect of superponderomotive electrons for laser driven ion acceleration. Realized by double-layer target configuration, we generated superponeromotive electrons from a first NCD density layer which drive consecutive ion acceleration from an attached nanometer thin foils. The significant enhancement of proton energies by a factor of 2.4 could represent a new path towards high efficient ion acceleration and also is of great importance of approaching relevant applications that require higher kinetic energies.
 
\begin{acknowledgments}
This work was supported by the DFG-funded Cluster of Excellence "Munich Centre for Advanced Photonics (MAP). The experiment was funded by EPSRC and the A-SAIL and LIBRA grants. We thank the staff of the Gemini operations team and the CLF for the assistance during the experiment. We also acknowledge essential target support from W.J. Ma. The code Epoch was in part funded by the UK EPSRC grants EP/G054950/1, EP/G056803/1, and EP/M022463/1.
\end{acknowledgments}

\bf
\end{document}